\documentclass[12pt,english]{revtex4}
\usepackage[T1]{fontenc}
\usepackage[latin1]{inputenc}

\makeatletter

\providecommand{\tabularnewline}{\\}

\usepackage{babel}
\makeatother
\newcommand{\ra}{\rightarrow}
\newcommand{\llb}{l^+ l^-}
\begin{document}

\title{New physics upper bound on the branching ratio of $B_{s}\rightarrow l^{+}l^{-}\gamma$.}

\author{Ashutosh Kumar Alok and S. Uma Sankar}

\address{Department of Physics, Indian Institute of Technology, Bombay, Mumbai
400076, India}

\begin{abstract}
We consider the effect of new physics on the branching ratio
of $B_{s}\rightarrow l^{+}l^{-}\gamma$ where $l=e,\,\mu$. If 
the new physics is of the form scalar/pseudoscalar, then it makes
{\it no} contribution to $B_s \rightarrow l^+ l^- \gamma$, unlike
in the case of $B_s \rightarrow l^+ l^-$, where it can potentially
make a very large contribution. If the new physics is in the form
of vector/axial-vector operators, then present data on 
$B \rightarrow (K,K^*) \llb$, does not allow a large enhancement for
$B(B_s \ra \llb \gamma)$. If the new physics is in the form of
tensor/pseudotensor operators, then the data on $B \ra (K,K^*) \llb$
gives no useful constraint but the data on $B \ra K^* \gamma$ does. 
Here again, a large enhancement of $B(B_s \rightarrow l^+ l^- \gamma)$,
much beyond the Standard Model expectation, is not possible. Hence, 
we conclude that the present data on $b \ra s$ transitions
allow a large boost in $B(B_s \ra \llb)$ but not in $B(B_s \ra \llb
\gamma)$.
\end{abstract}
\maketitle

The quark level transition $b \ra s \llb$ can lead to a number of 
important flavour changing neutral current (FCNC) decays in B mesons.
Among them are the semi-leptonic modes $B \ra (K, K^*) \llb$, the 
purely leptonic mode $B_s \ra \llb$ and the leptonic radiative mode 
$B_s \ra \llb \gamma$. The relationship between the semi-leptonic
and purely leptonic modes was discussed in \cite{alok-05}. It was shown
that, if the new physics occurs in the form of vector/axial-vector
operators, then present data on semi-leptonic branching ratios \cite{babar-03,
belle-03} constrain the branching ratio for $B_s \ra \llb$ to be of the
same order of magnitude as that of the Standard Model (SM). On the other
hand, if the new physics operators are in the form of scalar/pseudoscalar,
then the semi-leptonic branching ratios do not lead to any useful constraint 
on the rate for the purely leptonic mode. Hence a large enhancement of
$B_s \ra \llb$ is possible only if the new physics is in the form of 
scalar/psuedoscalar operators. Tensor/pseudotensor operators do not
contribute to $B_s \ra \llb$. In this letter, we examine the relation
between the rates for effective $b \ra s$ transitions and $B_s \ra
\llb \gamma$.

In the SM, the decay $B_{s}\rightarrow l^{+}l^{-}$ has small branching
ratio due to helicity suppression. The radiative decay 
$B_{s}\rightarrow l^{+}l^{-}\gamma$ is free from helicity suppression due
to emission of a photon in addition to the lepton pair. Thus the branching 
ratio for this leptonic radiative mode is much higher than that for the
purely leptonic mode despite an additional factor of $\alpha$. 
Because of this higher rate, this mode will be an important probe of 
$b \ra s \llb$ transitions which will be studied at present and future
experiments. The decays $B_{s}\rightarrow l^{+}l^{-}\gamma$ have
been studied in several papers \cite{eilam-97,aliev-97,geng-00,dincer-01,
kruger-03,scotes-03,melikhov-04} within the framework of SM. The 
effective new physics Lagrangian for $b \ra s \llb$ transition is the 
sum of three terms: vector/axialvector, scalar/pseudoscalar and 
tensor/pseudotensor. The
first two terms can arise both via penguin and box diagrams but the
last term arises only via the penguin diagram for $b \ra s \gamma$, 
in which the real photon is replaced by a virtual photon coupling to 
a lepton-antilepton pair.
In \cite{aliev-97,geng-00}, the effective $b \ra s \llb$ interaction
was dressed with an on-shell photon in all possible
ways. Helicity suppression is operative for the case where the photon
is emitted from the final lepton and the resultant amplitude is proportional 
to the lepton mass and is negligible. For the case where the photon is 
emitted from the internal lines of the $b \ra s$ loop transition, the 
amplitude is suppressed by factors $m_b^2/m_W^2$ and is also negligible. 
The main contribution to the $B_s \ra \llb \gamma$ amplitude comes from
the diagrams where the final state photon is emitted from either 
$b$ or $s$ quark in the effective $b \ra s \llb$ interaction. With this
procedure, the SM prediction for $B\left(B_{s}\rightarrow e^{+}e^{-}
\gamma\right)$ is calculated, in \cite{aliev-97,geng-00}, to be   
about $(2-7)\times10^{-9}$, with the rate for 
$B_{s}\rightarrow\mu^{+}\mu^{-}\gamma$ being a little lower.

In ref.\cite{dincer-01} a  higher value of branching ratio for
$B_s \ra \llb \gamma$ is predicted within SM. This higher value 
is due to a different parametrization of the form factors
$f_{V}$, $f_{A}$, $f_{TV}$ and $f_{TV}$. The parametrization of form
factors in \cite{aliev-97} is based on QCD sum rules whereas in
\cite{geng-00} it is based on light front models. But ref.\cite{dincer-01} 
uses the parametrization based on perturbative
QCD methods combined with heavy quark effective theory \cite{pirjol-00}.
In ref.\cite{melikhov-04}, it was argued that there are additional
contributions to the $B_s \ra \llb \gamma$ amplitude. The most important
one comes from the case where the real photon is emitted from the 
$b \ra s$ loop transition and the virtual photon, which pair produces 
the leptons, is emitted from the initial quarks. Due to this additional
amplitude, the SM prediction for $B\left(B_{s}\rightarrow e^{+}e^{-}
\gamma\right)$ in \cite{melikhov-04}, is about $2\times10^{-8}$, 
with the branching ratio for $B_{s}\rightarrow\mu^{+}\mu^{-}\gamma$,
being a little smaller compared to $B_{s}\rightarrow e^{+}e^{-}\gamma$.

In the present calculation, we are interested on how the current
data on $b \ra s$ transitions, due to the effective interactions
$b \ra s \llb$ and $b \ra s \gamma$, constrain the new physics
contribution to the leptonic radiative decays $B_s \ra \llb \gamma$. 

As mentioned earlier, new physics in the form of scalar/pseudoscalar
operators can give a large enhancement to the leptonic decay mode
$B_s \ra \llb$. The question then follows: What is the effect of these
operators on the leptonic radiative modes $B_s \ra \llb \gamma$? 
Unfortunately, scalar/pseudoscalar operators do not contribute to
$B_s \ra \llb \gamma$. The photon has $J=1$. Hence the $\llb$ pair
also must be in $J=1$ state so that the angular momentum of the final
state can be zero. However, by Wigner-Eckert theorem, the matrix element 
$\langle \llb (J=1) | \bar{l} (g_s + g_p \gamma_5) l | 0 \rangle$
is zero. This result also follows from direct calculation, as we 
illustrate below. 

We parametrize the scalar/pseudoscalar operator for 
$b\rightarrow sl^{+}l^{-}$ transition as
\begin{equation}
L_{SP}(b\rightarrow sl^{+}l^{-})=
\frac{G_{F}}{\sqrt{2}}\left(\frac{\alpha}{4\pi s_W^2}\right)
\bar{s}(g_{S}+g_{P}\gamma_{5})b \; \bar{l}(g_{S}^{'}+g_{P}^{'}\gamma_{5})l.
\end{equation}
The matrix element for $B_{s}\rightarrow l^{+}l^{-}\gamma$ is given by
\begin{equation}
M(B\rightarrow l^{+}l^{-}\gamma)=\frac{G_{F}}{\sqrt{2}}
\left(\frac{\alpha}{4\pi s_W^2}\right)
\left[g_{S}\langle\gamma\left|\overline{s}b\right|B_{s}(p)\rangle +
g_{P}\langle\gamma\left|\overline{s}\gamma_{5}b\right|B_{s}(p)\rangle\right]
\bar{u}(p_{l}) (g_S^{'} + g_P^{'} \gamma_5) v(p_{\bar{l}}).  
\end{equation}
To calculate the matrix elements of the quark operators
in the above equation, we need to first consider the following
vector and axial-vector matrix elements \cite{dincer-01,kruger-03},
\begin{eqnarray}
\langle \gamma(k)\left|\overline{s}\gamma_\mu b\right|B_{s}(p)\rangle
& = & e\epsilon_{\mu\nu\rho\sigma}\epsilon^{*\nu}p^{\rho}k^{\sigma}
f_{V}(q^{2})/m_{B_{s}}, \nonumber \\
\langle \gamma(k)\left|\overline{s}\gamma_{\mu}
\gamma_{5}b\right|B_{s}(p)\rangle 
& =& -ie\left[\epsilon_{\mu}^{*}(p\cdot k) - (\epsilon^{*}\cdot p)
k_{\mu}\right]f_{A}(q^{2})/m_{B_{s}},
\label{vme}
\end{eqnarray}
where $q=p_{l}+p_{\overline{l}}$.
Dotting the above equations with the momentum of $B_s$ meson $p^{\mu}$,
we get the scalar and pseudoscalar matrix elements to be identically zero,
\begin{equation}
\langle\gamma(k)\left|\overline{s}b\right|B_{s}(p)\rangle = 0 = 
\langle\gamma(k)\left|\overline{s}\gamma_{5}b\right|B_{s}(p)\rangle.
\end{equation}
That this amplitude vanishes, was also demonstrated in \cite{gaursrc}.
So, even if a large enhancement of $B_s \ra \llb$ is observed 
at LHC-b \cite{lhcb} due to new physics operators in scalar/psuedoscalar
form, there will be no corresponding enhancement of $B_s \ra \llb \gamma$. 

A legitimate question to ask at this stage is: Is it possible to have
a large enhancement of $B_s \ra \llb \gamma$ for any type of new physics
operator? Here we consider vector/axial-vector operators and 
tensor/pseudo-tensor operators one at a time and examine their contribution
to $B_s \ra \llb \gamma$ given the current experimental results on the 
$b \ra s$ transitions.

First we will assume that the new physics Lagrangian contains only
vector and axial-vector couplings. We parametrize it as
\begin{equation}
L_{VA}(b\rightarrow sl^{+}l^{-})=
\frac{G_{F}}{\sqrt{2}}\left(\frac{\alpha}{4\pi s_W^2}\right)
\bar{s}(g_{V}+g_{A}\gamma_{5})\gamma_{\mu}b \;
\bar{l}(g_{V}^{'}+g_{A}^{'}\gamma_{5})\gamma^{\mu}l,
\label{effL}
\end{equation}
where $g$ and $g^{'}$ are effective couplings which characterise
the new physics. 

The vector and axial-vector matrix elements are shown in 
Eq.~(\ref{vme}) \cite{dincer-01,kruger-03}. 
The $q^{2}$ dependence of the formfactors is parametrized 
as \cite{kruger-03,melikhov-04},
\begin{equation}
f_{i}(q^{2})=\beta_{i}\frac{f_{B_{s}}m_{B_{s}}}{
\Delta_{i}+0.5 m_{B_s}
\left(1-q^2/m_{B_s}^2\right)},
\label{ff}
\end{equation}
where $i=V,A,TA,TV$ and the parameters $\beta$ and $\Delta$ are
given in Table~I. 
\begin{table}[htb]
\caption{Parameters for the form factors}
\begin{tabular}{|c|c|c|c|c|}
\hline 
Parameter&
$f_{V}$&
$f_{TV}$&
$f_{A}$&
$f_{TA}$\tabularnewline
\hline
\hline 
$\beta(GeV^{-1})$&
0.28&
0.30&
0.26&
0.33\tabularnewline
\hline 
$\Delta(GeV)$&
0.04&
0.04&
0.30&
0.30\tabularnewline
\hline
\end{tabular}
\end{table}

The calculation of decay rate gives
\begin{equation}
\Gamma_{NP}\left(B_{s}\rightarrow l^{+}l^{-}\gamma\right)=
\left(\frac{G_{F}^{2}\alpha^{3}m_{B_{s}}^{5}f_{B_{s}}^{2}
}{3072\pi^{4}s_W^4}\right)
\left[g_{V}^{2}\left(g_{V}^{'2}+g_{A}^{'2}\right)\beta_{V}^{2}I_{V}+
g_{A}^{2}\left(g_{V}^{'2}+g_{A}^{'2}\right)\beta_{A}^{2}I_{A}\right],
\label{dr_va}
\end{equation}
where $I_{i}$ $(i=V,A)$ are the integrals over the dilepton invariant
mass $(z=q^{2}/m_{B_{s}}^{2})$. They are given by 
\begin{equation}
I_{i}=\int_{0}^{1}dz\,\frac{z(1-z)^{3}}{\left[
(\Delta_{i}/m_{B_s})+0.5(1-z)\right]^{2}}
\end{equation}
Here we have neglected the lepton masses in comparison to $m_{B}$
as we are only considering $l=e,\mu$. We will work under this approximation
throughout the paper.

In order to put bounds on $B_{NP}\left(B_{s}\rightarrow l^{+}l^{-}\gamma\right)$
we need to know the values of $g_{V}^{2}\left(g_{V}^{'2}+g_{A}^{'2}\right)$
and $g_{A}^{2}\left(g_{V}^{'2}+g_{A}^{'2}\right)$. For this we will
have to consider the semi-leptonic decay modes 
$B\rightarrow(K,\, K^{*})l^{+}l^{-}$.
The values of these quantities were calculated in \cite{alok-05},
\begin{eqnarray}
g_{V}^{2}(g_{V}^{'2}+g_{A}^{'2}) & = & (1.36_{-0.44}^{+0.53})\times10^{-2}
\nonumber \\
g_{A}^{2}(g_{V}^{'2}+g_{A}^{'2}) & = & (6.76_{-3.48}^{+4.04})\times10^{-3}.
\end{eqnarray}
These values were calculated under the assumption that 
$B_{NP}\left[B\rightarrow(K,\, K^{*})l^{+}l^{-}\right]=
B_{Exp}\left[B\rightarrow(K,\, K^{*})l^{+}l^{-}\right]$
i.e. the experimentally measured semi-leptonic  branching ratios 
are saturated by the new physics couplings. 
Putting these values in Eq.~(\ref{dr_va}), we get
\begin{equation}
B_{NP}\left(B_{s}\rightarrow l^{+}l^{-}\gamma\right)=
2.06_{-0.76}^{+0.84}\times10^{-9}.
\end{equation}
Therefore the upper bounds on the branching ratios are,
\begin{eqnarray}
B_{NP}\left(B_{s}\rightarrow l^{+}l^{-}\gamma\right) & \leq & 
2.90\times10^{-9} ~ {\rm at~ 1\sigma} \nonumber \\
B_{NP}\left(B_{s}\rightarrow l^{+}l^{-}\gamma\right) & \leq & 
4.58\times10^{-9} ~ {\rm at~ 3\sigma}.
\end{eqnarray}

These values are of the same order of magnitude as SM prediction. Thus 
we see that
we can't boost $B_{NP}\left(B_{s}\rightarrow l^{+}l^{-}\gamma\right)$
above its SM prediction even after assuming that the contribution
to the decay rate is totally due to new physics. 
The fact, that the experimentally measured values of the semileptonic 
branching ratios $B(B\rightarrow(K,\, K^{*})l^{+}l^{-})$
are close to their SM predictions, doesn't allow 
$B_{NP}\left(B_{s}\rightarrow l^{+}l^{-}\gamma\right)$
to have a value much different from its SM predictions if the new
physics responsible for this decay is of the form vector/axial-vector.
A more stringent upper bound is obtained if we equate the new physics
branching ratio to be the difference between the experimental value
and the SM prediction. In fact, this upper bound is consistent with
zero at $1~\sigma$ and is $B_{NP} (B_s \ra \llb \gamma) \leq 
2 \times 10^{-9}$ at $3~\sigma$.
Thus we can't boost $B_{NP}\left(B_{s}\rightarrow l^{+}l^{-}\gamma\right)$
much above its SM prediction if new physics is of the form vector/axial-vector.

We now consider new physics interaction in the form of tensors/pseudo-tensor
operators. We parametrize this Lagrangaian as,
\begin{equation}
L_{T}(b\rightarrow sl^{+}l^{-})=\frac{G_{F}}{\sqrt{2}}
\left(\frac{\alpha}{4\pi s_W^2}\right)\left(\frac{im_{b}}{q^{2}}\right)
\bar{s}\sigma_{\mu\nu}q^{\nu}(g_{TV}+g_{TA}\gamma_{5})b \; 
\bar{l}\gamma^{\mu}l.
\end{equation}
The necessary matrix element for $B_{s}\rightarrow l^{+}l^{-}\gamma$
is given by,
\begin{eqnarray}
\left\langle \gamma(k)\left|\overline{s}i\sigma_{\mu\nu}q^{\nu}b
\right|B_{s}(p)\right\rangle & = & -e\epsilon_{\mu\nu\rho\sigma}
\epsilon^{*\nu}p^{\rho}k^{\sigma}f_{TV}(q^{2}), \nonumber \\
\left\langle \gamma(k)\left|\overline{s}i\sigma_{\mu\nu}\gamma_{5}q^{\nu}b
\right|B_{s}(p)\right\rangle & = & -ie\left[\epsilon_{\mu}^{*}(p\cdot k)
-(\epsilon^{*}\cdot p) k_{\mu}\right]f_{TA}(q^{2}).
\label{tme}
\end{eqnarray}
The $q^{2}$ dependence of the formfactors is given in Eq.~(\ref{ff}).
The calculation of decay rate gives,
\begin{equation}
\Gamma_{NP}\left(B_{s}\rightarrow l^{+}l^{-}\gamma\right)=
\left(\frac{G_{F}^{2}\alpha^{3}m_{B_{s}}^{5}f_{B_{s}}^{2}}{
3072\pi^{4}s_W^4}\right)
\left[g_{TV}^{2}\beta_{TV}^{2}I_{TV} +
g_{TA}^{2}\beta_{TA}^{2}I_{TA}\right],
\label{dva_gamma}
\end{equation}
where $I_{i}$ $(i=TV,TA)$ are the integrals over the dilepton invariant
mass $(z=q^{2}/m_{B_{s}}^{2})$. They are given by 
\begin{equation}
I_{i}=\int_{(4m_{l}^{2}/m_{B_{s}^{2}})}^{1}dz\,
\frac{(1-z)^{3}}{z\left[(\Delta_{i}/m_{B_s}) + 
0.5 (1-z)\right]^{2}}.
\end{equation}
Here again we have neglected the lepton masses everywhere except in
the lower limit of the integral $I_{i}$ due to the presence of term
$1/z$ in the integrand. Thus we expect larger value for 
$\Gamma_{NP}\left(B_{s}\rightarrow e^{+}e^{-}\gamma\right)$ in comparison
to $\Gamma_{NP}\left(B_{s}\rightarrow\mu^{+}\mu^{-}\gamma\right)$
due to presence of term $lnz$ in the expression of decay rate.
We need to know the values of $g_{TV}^{2}$ and $g_{TA}^{2}$ in order to 
obtain the upper bound on $B_{NP} \left(B_{s}\rightarrow l^{+}l^{-}
\gamma\right)$. For this we will consider first the semi-leptonic 
decays $B\rightarrow(K,\, K^{*})l^{+}l^{-}$ and then the radiative
decay $B \ra K^* \gamma$.

In order to obtain bounds on $g_{TV}^{2}$, we will
have to consider the process $B\rightarrow Kl^{+}l^{-}$. The necessary
matrix element in this case is \cite{jaus-90,deshpande-88} ,
\begin{equation}
\left\langle K(p_{k})\left|\overline{s}i\sigma_{\mu\nu}q^{\nu}b
\right|B(p_{B})\right\rangle = \frac{1}{\left(m_{B}+m_{K^{*}}\right)}
q^{2}(p_{B}+p_{k})_{\mu}f_{T}(q^{2}).
\end{equation}
In above equation we have dropped a term proportional to $q_{\mu}$
as it will give rise to a term proportional to $(m_{l}/m_{B})^2$ in the
decay rate. The $q^{2}$ dependence of the formfactor is assumed to be
\begin{equation}
f_{T}(q^{2})=\frac{f_{T}(0)}{\left(1-q^2/m_B^2\right)}.
\end{equation}
The calculation of decay rate gives,
\begin{equation}
\Gamma_{NP}(B\rightarrow Kl^{+}l^{-}) =g_{TV}^{2}
\left(\frac{G_{F}^{2}m_{B}^{5}}{192\pi^{3}}\right)
\left(\frac{\alpha}{4\pi s_W^2}\right)^{2} f_{T}^{2}(0) I^{BK},
\label{dr_bk}
\end{equation}
where $I_{BK}$ is the integral over the dilepton invariant mass 
$(z=q^{2}/m_{B_{s}}^{2})$. This integral is given by 
\begin{equation}
I^{BK} = \int_{z_{min}}^{z_{max}}dz\,
\frac{\phi(z)^{3/2}}{2(1+k)^{2}(1-z)^{2}},
\end{equation}
where $\phi(z)=\left(z-1-k^{2}\right)^{2}-4k^{2}$ with $k=m_{K}/m_{B}$
and the limits of integration for $z$ are given by 
$z_{min}=4m_{l}^{2}/m_{B}^{2}$ and $z_{max}=(1-k)^{2}$.

Here again we make the approximation $\Gamma_{NP}=\Gamma_{Exp}$ .
Under this approximation we get from Eq.~(\ref{dr_bk}),
\begin{equation}
g_{TV}^{2} = \frac{B_{Exp}(B\rightarrow Kl^{+}l^{-})}{
2.35\left[f^{+}(0)\right]^{2}}\times10^{4}.
\end{equation}
In order to obtain bounds on $g_{TA}^{2}$, we will
have to consider the process $B\rightarrow K^{*}l^{+}l^{-}$. The
necessary matrix elements in this case are \cite{jaus-90,deshpande-88},
\begin{eqnarray}
\left\langle K^{*}(p_{k})\left|\overline{s}i\sigma_{\mu\nu}
q^{\nu}b\right|B(p_{B})\right\rangle & = & iT_{1}(q^{2})
\epsilon_{\mu\nu\rho\sigma}\epsilon^{*\nu}
(p_{B}+p_{k})^{\rho}(p_{B}-p_{k})^{\sigma}, \nonumber \\
\left\langle K^{*}(p_{k})\left|\overline{s}i\sigma_{\mu\nu}
q^{\nu}\gamma_{5}b\right|B(p_{B})\right\rangle & = & T_{2}(q^{2})
(m_{B}^{2}-m_{K^{*}}^{2})\epsilon_{\mu}^{*}+T_{3}(q^{2})
(\epsilon^{*}\cdot p_{B})(p_{B}+p_{K^{*}})_{\mu}.
\label{tmesld}
\end{eqnarray}
Here again we have dropped the terms proportional to $q_{\mu}$. The
$q^{2}$dependence of the formfactors is assumed to have the from,
\begin{equation}
T_{i}(q^{2})=\frac{T_{i}(0)}{\left(1-q^2/m_B^2\right)},
\end{equation}
where $i=1,2,3.$

The calculation of decay rate gives, 
\begin{equation}
\Gamma_{NP}(B\rightarrow K^{*}l^{+}l^{-}) = 
\left(\frac{G_{F}^{2}m_{B}^{5}}{192\pi^{3}}\right)
\left(\frac{\alpha}{4\pi s_W^2}\right)^{2}
\left[g_{TV}^{2}T_{1}^{2}(0)I_{TV}^{BK^{*}} +
g_{TA}^{2}T_{2}^{2}(0)I_{TA}^{BK^{*}}\right],
\label{dr_bk*}
\end{equation}
where $I_{i}^{BK^{*}}$ $(i=TV,TA)$ are the integrals over the dilepton
invariant mass $(z=q^{2}/m_{B_{s}}^{2})$. They are given by 
\begin{eqnarray}
I_{TV}^{BK^{*}} & = & \int_{z_{min}}^{z_{max}}\frac{dz}{z(1-z)^{2}}\,
\phi(z)^{3/2} \nonumber \\
I_{TA}^{BK^{*}} & = & \int_{z_{min}}^{z_{max}}\frac{dz}{z^{2}(1-z)^{2}}\,
\phi(z)^{3/2} \left[\frac{(1-k^{*2})^{2}}{2\phi(z)} 
\left\{ 2z+\frac{(1-k^{*2}-z)^{2}}{4k^{*2}}\right\} \right. \nonumber \\
& & \left. + \frac{1}{8k^{*2}} +
\frac{(1-k^{*2})(1-z-k^{*2})}{4k^{*2}}\right]
\label{IBK*A}
\end{eqnarray}
with $k^{*}=\frac{m_{K^{*}}}{m_{B}}$. Here we assumed 
$T_{2}(0)\simeq T_{3}(0)$. $\phi(z)$ and $z_{max}$ are the same as 
in the case of $B\rightarrow Kl^{+}l^{-}$ with $k$ replaced by $k^{*}$.
For $B\rightarrow K^{*}l^{+}l^{-}$, the experimental branching ratio 
is given with a lower cut on the di-lepton invariant mass, $m_{l\bar{l}}
>0.14$ GeV, in order to supress background from photon conversions and 
$\pi^{0}\rightarrow e^{+}e^{-}\gamma$ \cite{belle-03}. 
We use this cut as the lower limit of integration for $z$.
In all previous kinematic integrals, $z_{min}$, the lower limit of intergration
for $z = q^2/m_B^2$ is taken to be the theoretical minimum $4 m_l^2/m_B^2$.
The kinematic integral, $I_{TA}^{BK^*}$ in Eq.~(\ref{IBK*A}), contains
a $1/z^2$ term which comes from the propagator of virtual
photon pair producing a lepton anti-lepton pair. At very small values of
$z$, this term dominates the integral and makes it very large. However,
experimentally the lower limit on $q^2$ is much larger than the theoretical
lower limit. Therefore, in calculating the bounds on new physics, the 
lower limit of $q^2$ in the theoretical calculation should be the same
as the experimental lower limit. 

Under the assumption $\Gamma_{NP} =
\Gamma_{Exp}$ and using Eq.~(\ref{dr_bk*}), we get
\begin{equation}
g_{TA}^{2} = \frac{B_{Exp}(B\rightarrow K^{*}l^{+}l^{-})\times10^{3} 
-1.37I_{TV}^{BK^{*}}T_{1}^{2}(0)g_{TV}^{2}}{1.37I_{TA}^{BK^{*}}T_{2}^{2}(0)}.
\end{equation}
In our calulation we take the formfactors to be \cite{ali-00},
$f_{T}(0)=0.355_{-0.055}^{+0.016}$, $T_{1}(0)=0.379_{-0.045}^{+0.058}$,
$T_{2}(0)=0.379_{-0.045}^{+0.058}$.
The experimentally measured values of the branching ratios are
\cite{belle-03},
$B_{Exp}(B\rightarrow Kl^{+}l^{-}) = 
(4.8_{-0.9}^{+1.0}\pm0.3\pm0.1)\times10^{-7}$ and
$B_{Exp}(B\rightarrow K^{*}l^{+}l^{-}) = 
(11.5_{-2.4}^{+2.6}\pm0.8\pm0.2)\times10^{-7}$.
Adding all errors in quadrature, we get
$g_{TV}^{2}=1.63_{-0.60}^{+0.39}\times10^{-2}$
for $l=e,\mu$. 
The best fit values for $g_{TA}^2$ turn out to be negative
and very small (${\cal O} \simeq 10^{-6}$). The fact that these come out 
to be negative means that the semi-leptonic decay rates 
{\it can not} be explained
purely in terms of tensor/pseudo-tensor operators. Imposing the 
condition that $g_{TV}^2$ and $g_{TA}^2$ should be non-negative, gives
us the conditions 
\begin{equation}
g_{TA}^2 = 0,~{\rm and}~  
g_{TV}^{2}=1.63_{-0.60}^{+0.39}\times10^{-2}~{\rm for}~ l = e, \mu.
\label{gtvgta}
\end{equation}

The branching ratio for $B_{s}\rightarrow l^{+}l^{-}\gamma$, due
to $L_{T}$ is,
\begin{equation}
B_{NP}\left(B_{s}\rightarrow l^{+}l^{-}\gamma\right) = 
\left[3.15 \, I_{TV} g_{TV}^{2} + 3.81 \, I_{TA} g_{TA}^{2} \right]
f_{B_{s}}^{2}\times10^{-6}.
\label{bt_llgamma}
\end{equation}
Substituting $f_{B_{s}}=240\pm30$ MeV \cite{lattice} and 
the values of $g_{TV}^2$ and $g_{TA}^{2}$ in 
Eq.~(\ref{bt_llgamma}), we get
\begin{eqnarray}
B_{NP}\left(B_{s}\rightarrow e^{+}e^{-}\gamma\right) & = & 
1.91_{-0.85}^{+0.66}\times10^{-7} \nonumber \\
B_{NP}\left(B_{s}\rightarrow\mu^{+}\mu^{-}\gamma\right) & = & 
6.45_{-2.86}^{+2.24}\times10^{-8}.
\end{eqnarray}
Therefore the upper bounds on the branching ratios are,
\begin{eqnarray}
B_{NP}\left(B_{s}\rightarrow e^{+}e^{-}\gamma\right) & \leq & 
2.57\times10^{-7}, \nonumber \\
B_{NP}\left(B_{s}\rightarrow\mu^{+}\mu^{-}\gamma\right) & \leq & 
8.69\times10^{-8}
\label{nptensor1}
\end{eqnarray}
at $1\sigma$ and
\begin{eqnarray}
B_{NP}\left(B_{s}\rightarrow e^{+}e^{-}\gamma\right) & \leq & 
3.89\times10^{-7}, \nonumber \\
B_{NP}\left(B_{s}\rightarrow\mu^{+}\mu^{-}\gamma\right) & \leq & 
1.32\times10^{-7}
\label{nptensor3}
\end{eqnarray}
at $3\sigma$.
These branching ratios are about 40-50 times greater than
the predictions in \cite{eilam-97,geng-00}. Thus the data on semileptonic
decays allows an enhancement of one to two orders of magnitude in 
$B_{NP}\left(B_{s}\rightarrow l^{+}l^{-}\gamma\right)$
if new physics interactions are of type tensor/pseudo-tensor. 

Here we note that $b \ra s \gamma$ transition also has a tensor 
operator and we consider the constraint on the tensor/pseudotensor
contribution to $B_s \ra \llb \gamma$ from the experimentally 
measured value of the branching ratio of $B\rightarrow K ^{*}\gamma$.
For this we consider the quark level interction $b\rightarrow s\gamma$.
We parametrize new physics effective Lagrangian for $b\rightarrow s\gamma$
as
\begin{equation}
L(b\rightarrow s\gamma)=
\left(\frac{G_{F}}{\sqrt{2}}\right)
\left(\frac{iem_{B}}{16\pi^{2}s_W^2}\right)
\overline{s}\sigma_{\mu\nu}q^\nu(g_{TV}+g_{TA}\gamma_{5})b\, 
\epsilon^{(\gamma)\mu}.
\end{equation}
where $\epsilon^{(\gamma) \mu}$ is the polarization vector of
the photon and $q^\nu$ is its momentum. 
Replacing $\epsilon^{(\gamma) \mu}$ by $(e/q^2) \bar{l} \gamma^\mu l$
gives rise to $b \ra s \llb$ tensor/pseudotensor operators. Thus 
the present experimental limit on $B \ra K^* \gamma$ leads 
to a bound on $B_s \ra \llb \gamma$ arising from new physics 
operators of tensor/pseudotensor form. 

The amplitude for $B\rightarrow K^{*}\gamma$ is given by,
\begin{equation}
A(B\rightarrow K^{*}\gamma)=
\left(\frac{G_{F}}{\sqrt{2}}\right)
\left(\frac{em_{B}}{16\pi^{2}s_W^2}\right)
\epsilon^{(\gamma)\mu}\left\langle K^{*}(p_{k})
\left|\overline{s}\sigma_{\mu\nu}q^{\nu}(g_{TV}+
g_{TA}\gamma_{5})b\right|B(p)\right\rangle. 
\end{equation}
The necessary matrix elements are given in Eq.~(\ref{tmesld}) 
with $T_{1}(0)=T_{2}(0)$ for real photon emission.
The calculation of decay rate gives,
\begin{equation}
\Gamma_{NP}(B\rightarrow K^{*}\gamma)=
(g_{TV}^{2}+g_{TA}^{2})\left(\frac{G_{F}^{2}\alpha}{1024\pi^{4}
s_W^4}\right)m_{B}^{5}(1-k^{*2})^{3}T_{1}^{2}(0).
\end{equation}

The process $B\rightarrow K^{*}\gamma$ has been observed with a branching
ratio \cite{babar-04},
\begin{equation}
B_{Exp}(B\rightarrow K^{*}\gamma)=(3.92\pm0.20\pm0.24)\times10^{-5}.
\end{equation}
Under the assumption $\Gamma_{NP}(B\rightarrow K^{*}\gamma)
=\Gamma_{Exp}(B\rightarrow K^{*}\gamma)$,
we get
\begin{equation}
g_{TV}^{2}+g_{TA}^{2}=1.92_{-0.48}^{+0.59}\times10^{-4}.
\end{equation}
Comparing this constraint with those in Eq.~(\ref{gtvgta}) we see that 
the process $B\rightarrow K^{*}\gamma$ puts a much stronger constraint 
on the $g_{TV}^2$ in comparison to that from
$B\rightarrow(K,K^{*})l^{+}l^{-}$. We substitute the above limit in
Eq.~(\ref{dva_gamma}) along with the approximation $\beta_{TV} \simeq 
\beta_{TA} = 0.33$ GeV$^{-1}$. The phase space integrals $I_{TV}$ and
$I_{TA}$ are essentially equal to each other. For electrons their
value is $64$ and for muons their value is $22$. Then we get  
the branching ratios to be
\begin{eqnarray}
B_{NP}(B_{s}\rightarrow e^{+}e^{-}\gamma) & = & 
2.71_{-0.95}^{+1.10}\times10^{-9}, \nonumber \\
B_{NP}(B_{s}\rightarrow\mu^{+}\mu^{-}\gamma) & = & 
9.18_{-3.25}^{+3.64}\times10^{-10}.
\end{eqnarray}
These values are of the same order as SM predictions. Thus the stronger
constraint on tensor/pseudo-tensor couplings coming
from the experimentally measured value of $B(B\rightarrow K^{*}\gamma)$
doesn't allow an enhancement of $B_{NP}(B_{s}\rightarrow l^{+}l^{-}\gamma)$.

\emph{Conclusions}\textbf{\emph{.}} 
The quark level interaction $b\rightarrow sl^{+}l^{-}$
is responsible for the three types of decays
(a) semi-leptonic $B \ra (K,K^*) \llb$, 
(b) purely leptonic $B_{s}\rightarrow l^{+}l^{-}$ and also
(c) leptonic radiative $B_{s}\rightarrow l^{+}l^{-}\gamma$. 
It was shown in previously \cite{alok-05} that if the purely leptonic
branching ratio $B(B_{s}\rightarrow\mu^{+}\mu^{-})$$\geq$
$10^{-8}$ then new physics operators responsible for this
have to be of the form scalar/pseudoscalar. 
Here we have shown that such scalar/pseudoscalar operators
have no effect on the leptonic radiative modes $B_s \ra \llb \gamma$.
Regarding other types of new physics operators, the vector/axial-vector
operators {\it can not} enhance the branching ratios of 
$B_s \ra \llb \gamma$
much beyond their SM values, given the constraints coming from
the measured semi-leptonic rates. 
New physics operators in the form of tensor/pseudo-tensor also 
{\it can not} enhance $B_s \ra \llb \gamma$ branching ratios 
given the constraints coming from $B \ra K^* \gamma$.
{\bf Thus we are led to the conclusion that the present 
data on $b \ra s$ transitions allow a large boost in 
$B(B_{s}\rightarrow l^{+}l^{-})$ but not   
in $B(B_{s}\rightarrow l^{+}l^{-}\gamma)$}.

\textbf{Acknowledgement}

This work grew out of discussions during the Ninth Workshop on High
Energy Physics Phenomenology (WHEPP9) at Bhubaneshwar, India. We thank
the members of the B-physics working group, especially Prof. David Hitlin, 
California Institute of Technology, USA, for valuable suggestions. We 
also thank Dr. Naveen Gaur, Delhi University, for an enlightening comment.

\end{document}